\documentclass[11pt]{article}

\usepackage{amssymb}
\usepackage[dvips]{graphicx}
\usepackage{epsfig}
\usepackage{cite}

\def\laq{~\raise 0.4ex\hbox{$<$}\kern -0.8em\lower 0.62
ex\hbox{$\sim$}~}
\def\gaq{~\raise 0.4ex\hbox{$>$}\kern -0.7em\lower 0.62
ex\hbox{$\sim$}~}

\def\be{\begin{equation}}
\def\ee{\end{equation}}
\def\bea{\begin{eqnarray}}
\def\eea{\end{eqnarray}}
\newcommand{\nn}{\nonumber}
\newcommand{\de}{\partial}

\def \ra {\rightarrow}
\def \l {\lambda}
\def \L {\Lambda}
\def \d {\delta}

\def \g {\gamma}
\def \s {\sigma}

\def \e {\epsilon}

\def \Om {\Omega}

\def \5 {{}^{(5)}}
\def \cd {\nabla}

\begin{document}
\begin{titlepage}

\begin{flushright}

\end{flushright}

\vspace*{1.5 cm}

\begin{center}

\Huge
{Bouncing cosmology from Kalb-Ramond Braneworld}

\vspace{1cm}

\large{G. De Risi$^{1,2}$}

\vspace{.2in}

\normalsize
{\sl $^1$Institute of Cosmology \& Gravitation,
University of Portsmouth, Portsmouth~PO1 2EG, UK}\\

\vspace{.2in}

{\sl $^2$Istituto Nazionale di Fisica Nucleare, 00186~Roma, Italy}

\vspace*{1.5cm}

\begin{abstract}

We consider a 3-brane embedded in a warped 5-dimensional background with a dilaton and a Kalb-Ramond 2-form.
We show that it is possible to find static solutions of the form of charged dS/AdS-like black hole with horizon
which could have a negative mass parameter. The motion of the 3-brane in this bulk generates an effective
4-dimensional bouncing cosmology induced by the negative dark radiation term. This model avoids the instability
that arises for bouncing brane in a Reissner-Nordstr{\o}m-AdS bulk.

\end{abstract}
\end{center}

\end{titlepage}
\newpage

\section{Introduction}

Braneworld models \cite{Randall:1999ee,Randall:1999vf} have generated, during the past decade,
enormous attention, due to the dramatic change they
inspired in our understanding of extra dimensions. According to this framework, our universe is a ``brane''
embedded in a higher-dimensional space, on which the Standard Model fields are confined, while gravity
is localized near the brane by the warped geometry of the extra dimension. It is possible to
construct models in which the brane evolution mimics a Friedmann-Robertson-Walker (FRW) cosmology,
with modifications at small scales due to the gravitational effect of the bulk spacetime on the brane \cite{Binetruy:1999ut,Binetruy:1999hy,Shiromizu:1999wj}.
In particular, provided the bulk is taken to be a Reissner-Nordstr{\o}m-AdS black hole, such modifications
can lead to bouncing 4D cosmological models \cite{Mukherji:2002ft}. Unfortunately the brane, during its evolution
in the bulk, always crosses the Cauchy horizon of the AdS black hole, which is unstable \cite{Kanti:2003bx,Hovdebo:2003ug}.


In this paper we present a different model, in which this problem is avoided. We consider
a brane embedded in a supergravity background in which both the dilaton and the Kalb-Ramond 2-form are
turned on (but without a dilaton potential). By dualizing the 2-form, we obtain Einstein-Maxwell like equations
of motion, but with a different sign for the kinetic term of the Maxwell-like field. The static solution is therefore
different, and the term that dominates at high curvature is like ``stiff matter'' with positive energy
density. Even though this implies that the energy contribution at high curvatures is positive, so that
it can not drive a bounce, this opens an interesting possibility of having negative energy contributions
at intermediate curvatures, by letting the mass of the black hole be negative.
The parameter space allows this while avoiding a naked singularity. In this case, we show that
it is possible that the brane bounces before crossing the black hole horizon, so that the effective
4-dimensional cosmological evolution will not suffer the instability of \cite{Hovdebo:2003ug}.

The paper is organized as follows: In section \ref{Bulk} we derive the bulk equations and present the
static solution, which is an asymptotically (A)dS black hole.
charged under the Kalb-Ramond field with the dilaton frozen at its VEV. Section \ref{Horizons}
is devoted to the study of the position of the horizons, and it is shown that in a certain region of the
parameter space (in particular for a negative cosmological constant) there is no naked singularity even
if the mass parameter of the black hole is negative. Then,
in section \ref{Branecosm} we embed the brane and allow it to move, to mimic a cosmological evolution for
a brane observer \cite{Kraus:1999it}. The occurrence of the bounce and its position is discussed here. Finally,
in section \ref{Concl} we present our conclusions and discuss open problems.

\section{The 5-D solution}
\label{Bulk}

Consider the low-energy string effective action:
\bea
S &=& \frac{M^3}{2} \int d^5 x \sqrt{- g } \left( R - 2\L e^{\s_1 \phi} - \frac{1}{12}H_{ABC}H^{ABC}e^{\s_2 \phi}
-g^{AB} \de_A \phi \de_B \phi \right) \nn \\
& & -  T_3 \int d^{4} \xi \sqrt{-\g} e^{\l \phi}
\label{Action}
\eea
(with $H_{ABC}=\de_{[A}B_{BC]}$), which describes a 3-brane embedded in a
5-dimensional bulk with dilaton and Kalb-Ramond 2-form \cite{Behrndt:1994ev,Copeland:1994vi}.
We set the dilaton potential to zero for simplicity, and allow the dilaton to be non-minimally coupled with
the cosmological constant, the antisymmetric tensor and the brane \cite{Maeda:2000wr,DeRisi:2006pz}.
However, the brane is taken to be neutral with respect to the antisymmetric field.
The presence of the brane action in (\ref{Action}) gives a singular part, which we will take into account
by the Israel junction condition \cite{Maartens:2003tw,Kraus:1999it}, so now we will consider only the bulk part.
The equations of motion are:
\bea
&& G_{AB} = \left[ -e^{\s_1 \phi}\L - \frac{1}{2} \left( \de_c \phi \right)^2 - \frac{1}{24} e^{\s_2 \phi} H^2
\right] g_{AB} + \de_A \phi \de_B \phi + \frac{1}{4} e^{\s_2 \phi} H_{ACE}H_B^{~~CE},
\label{einsteq} \\
&& \cd_A \cd^A \phi - \s_1  e^{\s_1 \phi} \L- \frac{\s_2}{24} e^{\s_2 \phi} H^2  = 0,
\label{dileq} \\
&& \cd_C \left( e^{\s_2 \phi}H^{CAB}\right) = 0,
\label{Heq}
\eea
where $H^2 = H_{ABC}H^{ABC}$.

The equation for the antisymmetric tensor (\ref{Heq}) can be solved by the ansatz\footnote{See also
\cite{Das:2005xk}, though our solutions differ from theirs.}
\be
H^{CAB} = \e^{CABDE}\cd_D A_E e^{-\s_2 \phi}.
\label{HAnsatz}
\ee
Substituting this into eq. (\ref{Heq}) we get
\bea
\cd_C \left( e^{\s_2 \phi}\e^{CABDE}\cd_D A_E e^{-\s_2 \phi}\right) =
\e^{ABCDE}\cd_C \cd_D A_E &=& \frac{1}{2}\e^{ABCDE}\left(\cd_C \cd_D - \cd_D \cd_C \right)A_E \nn \\
&=& \frac{1}{4}\e^{ABCDE}R_{CDE}^{~~~~~M}A_M = 0.
\eea
The last equality follows from the second Bianchi identity $R_{[ABC]}^{~~~~~~~D}=0$.
Now the equation for $A_M$ can be obtained by the identity $\de_{[A} H_{BCD]} = 0$, while we have to substitute
the expression we get for $H$ in eqs. (\ref{einsteq}) and (\ref{dileq}) to get the correct equations of motion.
Observing that
\bea
H^2 &=&  12 e^{-2 \s_2} F^2, \nn \\
H_{AMN}H_B^{~~MN} &=& 2 e^{-2 \s_2} \left( 2F^2g_{AB} -4F_{AC}F_B^{~~C} \right),
\label{HtoF}
\eea
where
\be
F_{MN} = \frac{1}{2} \left( \cd_M A_N - \cd_N A_M \right),
\label{Fdef}
\ee
we get the following equations:
\bea
&& G_{AB} = \left[ -e^{\s_1 \phi}\L - \frac{1}{2} \left( \de_c \phi \right)^2 + \frac{1}{2} e^{-\s_2 \phi} F^2
\right] g_{AB} + \de_A \phi \de_B \phi - 2 e^{-\s_2 \phi} F_{AC}F_B^{~~C},
\label{DualEinsteq} \\
&& \cd_A \cd^A \phi - \s_1  e^{\s_1 \phi}\L - \frac{\s_2}{2} e^{-\s_2 \phi} F^2 = 0,
\label{DualDileq} \\
&& \cd_B \left( e^{-\s_2 \phi}F^{AB}\right) = 0,
\label{DualFeq}
\eea
which look like the equations of motion of an Einstein-Maxwell model, but in this case the signs
of the Maxwell fields are reversed.

Now, following \cite{Kraus:1999it} we search for a static solution with a maximally symmetric 3-space.
The metric ansatz is:
\be
ds^2 = -f(R) dt^2 +R^2 \left( \frac{dr^2}{1-k r^2} + r^2 d\Om^2 \right) + \frac{dR^2}{f(R)}.
\label{metricansatz}
\ee
We also assume that the gauge field is purely electric, i.e. $A_M \equiv \left(A(R),{\bf 0},0 \right)$, and
that the dilaton also depends only on the radial coordinate, $\phi = \phi(R)$.
Then the Einstein equations read:
\bea
-\frac{3}{2} \frac{f'}{R} - 3 \frac{f}{R^2} + 3\frac{k}{R^2} &=& \L e^{\s_1 \phi} - \frac{1}{4} A'^2e^{-\s_2 \phi} +
\frac{1}{2} f \phi', \nn \\
\frac{3}{2} \frac{f'}{R}  + 3 \frac{f}{R^2} - 3\frac{k}{R^2} &=& -\L e^{\s_1 \phi}  + \frac{1}{4} A'^2e^{-\s_2 \phi} +
\frac{1}{2} f \phi', \nn \\
\frac{f''}{2} +2 \frac{f'}{R} + \frac{f}{R^2} - \frac{k}{R^2} &=& -\L e^{\s_1 \phi} - \frac{1}{4} A'^2e^{-\s_2 \phi}
- \frac{1}{2} f \phi'.
\label{Eeq1}
\eea
It is easy to see that, summing up the first and second equations, we get that the dilaton must be constant,
$\phi'(R)=0$, so that we can reabsorb it into the definition of the coupling constant. So, we are left with
only the vector field; the equations are
\bea
\frac{3}{2} \frac{f'}{R}  + 3 \frac{f}{R^2} - 3\frac{k}{R^2} &=& -\L + \frac{1}{4} A'^2, \nn \\
\frac{1}{2} f''+ 2 \frac{f'}{R}  +  \frac{f}{R^2} - \frac{k}{R^2} &=& -\L - \frac{1}{4} A'^2, \nn \\
A'' + 3\frac{A'}{R} &=& 0.
\label{compeq}
\eea
As usual, by the conservation of the energy-momentum tensor, only two of these equations are independent.
A solution of these equation can be cast in the form
\bea
f(R) &=& -\frac{Q^2}{3R^4} - \frac{\mu}{R^2} + k - \frac{\L}{6} R^2, \nn \\
A(R) &=& \pm \frac{Q}{R^2}.
\label{backgroundsol}
\eea
Notice that this solution is quite similar to the one that was found in \cite{Mukherji:2002ft}, but the term proportional
to the charge is opposite in sign. Because of this, it is possible to avoid a naked singularity even if $\mu$ is
negative. In the next section we will discuss the location of horizons for different values of the physical
constants.

\section{Horizons in the Kalb-Ramond Black Hole}
\label{Horizons}

We can track the location of the horizons by finding the zeros of the metric function $f$ in (\ref{backgroundsol}). For simplicity
we will assume the spatial part of the metric to be flat, $k = 0$
The zeroes of the function $f$ can be found by solving the equation:
\be
x^3 + 6\frac{\mu}{\L}x + 2\frac{Q^2}{\L} = 0,
\label{horeq}
\ee
with $x = R^2$ (so that we are interested only in positive solutions). This equation is conveniently solved by use of
the Chebyshev radicals. We can identify three different cases (let us stress that we will assume $\mu < 0$ from now on):
\begin{description}
\item[\underline{\textmd{Case 1}}] When
\be
\L > -\frac{8\mu^3}{Q^4},
\label{horint1}
\ee
we have one real solution
\be
x = -2 \sqrt{-\frac{2\mu}{\L}} \cosh \left[ \frac{1}{3}
\cosh^{-1} \left( \frac{Q^2\sqrt{\L}}{(-2\mu)^{3/2}}\right)\right],
\label{horsol1}
\ee
but this solution is always negative, so it is not acceptable. Therefore in this part of the
parameter space the background has a naked singularity.

\item[\underline{\textmd{Case 2}}] When
\be
 0< \L < -\frac{8\mu^3}{Q^4},
\label{horint2}
\ee
we have three real solutions:
\bea
x_1 &=& 2 \sqrt{-\frac{2\mu}{\L}} \cos \left[ \frac{1}{3}
\arccos \left( -\frac{Q^2\sqrt{\L}}{(-2\mu)^{3/2}}\right)\right], \nn \\
x_2 &=& -2 \sqrt{-\frac{2\mu}{\L}} \cos \left[ \frac{1}{3}
\arccos \left( \frac{Q^2\sqrt{\L}}{(-2\mu)^{3/2}}\right)\right], \nn \\
x_3 &=& -x_1-x_2,
\label{horsol2}
\eea
of which the second one is negative, $x_2<0$ thus unacceptable, and $x_1 > x_3$. In this case we have
two horizons, just as in the usual Reissner-Nordstr{\o}m black hole. The horizons can be written as:
\bea
R_+ &=& \sqrt{2} \left(-\frac{2\mu}{\L}\right)^{1/4} \sqrt{\cos \left[ \frac{1}{3}
\arccos \left( -\frac{Q^2\sqrt{\L}}{(-2\mu)^{3/2}}\right)\right]} \label{plushor}, \\
R_- &=& \sqrt{2} \left(-\frac{2\mu}{\L}\right)^{1/4} \sqrt{\cos \left[ \frac{1}{3}\left( \pi +
\arccos \left( \frac{Q^2\sqrt{\L}}{(-2\mu)^{3/2}}\right) \right) \right]} \label{minushor}.
\eea
\item[\underline{\textmd{Case 3}}] This case occurs when
\be
\L < 0,
\label{horint3}
\ee
and we have, again, one real solution, which can be written as
\bea
x &=& 2\sqrt{\frac{2\mu}{\L}} \cos \left[ \frac{1}{3} \arccos
\left( \frac{Q^2\sqrt{-\L}}{(-2\mu)^{3/2}}\right)\right] ~~~~~~{\rm for} ~~~~ \frac{8\mu^3}{Q^4} < \L < 0, \nn \\
x &=& 2\sqrt{\frac{2\mu}{\L}} \cosh \left[ \frac{1}{3}
\cosh^{-1} \left( \frac{Q^2\sqrt{-\L}}{(-2\mu)^{3/2}}\right)\right] ~~~~{\rm for} ~~~~\L < \frac{8\mu^3}{Q^4}.
\label{horsol3}
\eea
This solution is always positive, so that in this case we have only one horizon, which can be written as
\be
R_0 = \left(\frac{2\mu}{\L}\right)^{1/4} \sqrt{{\rm C_{1/3}} \left( \frac{2Q^2\sqrt{-\L}}{(-2\mu)^{3/2}}\right)},
\label{singlehor}
\ee
where the Chebyshev polynomial ${\rm C_{1/3}}$ is intended to be
\be
{\rm C_{1/3}}(t) \equiv \left\{\begin{array}{lc} 2\cos \left(\frac{1}{3}\arccos(\frac{t}{2})\right) & 0<t<2  \\
 2\cosh \left(\frac{1}{3}\cosh^{-1}(\frac{t}{2})\right) & t>2. \end{array} \right.
\label{Chebys}
\ee
Since the bulk is asymptotically AdS, this is the most interesting case. As we will see in the next section,
it is in this case that the cosmological evolution on the brane undergoes a viable bounce.

\end{description}

\section{Cosmological evolution of the moving brane }
\label{Branecosm}

In this section we embed a 3-brane in the bulk described previously, by cutting out $R > a$, and imposing
a $\mathbb{Z}_2$ symmetry at the edge. We then let the brane move through the bulk, $a=a(\tau)$,
where $\tau$ is proper time on the brane.
The movement of the brane in the 5D bulk induces a cosmological evolution on the brane
via the Israel junction condition \cite{Kraus:1999it}. To see this, we need to calculate the extrinsic curvature
of the brane. The unit vectors tangent and normal to the moving brane are:
\bea
u^A &=& \left( -\frac{\sqrt{f + \dot{a}^2}}{f},{\bf 0},\dot{a} \right);~~~~~~ v^A_{\ \ i} = \d^A_{\ \ i}, \nn \\
n_A &=& \left( \dot{a},{\bf 0},-\frac{\sqrt{f + \dot{a^2}}}{f} \right),
\label{vectorbasis}
\eea
so that the (spatial part of) the extrinsic curvature is
\be
K_{ij} = \frac{1}{2}\frac{\sqrt{f + \dot{a}^2}}{a}\g_{ij},
\label{extrcur}
\ee
where $\g_{\mu \nu}$ is the induced metric on the brane, which is FRW with scale factor $a$, and the dot represents
a derivative with respect to proper time on the brane. The modified Friedman equations are obtained from the
junction conditions,
\be
K_{ij} = \frac{1}{M^3} \left( T_{ij} - \frac{1}{3}T\g_{ij} \right).
\ee
where $T_{\mu \nu}$ is the brane energy-momentum tensor. For $T_{\mu \nu} = \lambda \g_{\mu \nu} + (\rho + p)u_\mu u_\nu
+ p\g_{\mu \nu}$, where $\l$ is the brane tension, we find that:
\be
H^2 = \frac{\rho}{3M_4^2} + \frac{\L_4}{3} - \frac{k}{a^2} + \left(\frac{\rho}{3M^3}\right)^2
+ \frac{Q^2}{3a^6} + \frac{\mu}{a^4},
\label{complFried}
\ee
where
\be
M_4 = \frac{2}{3} \frac{\l}{M^6}, \qquad \L_4 = \frac{1}{3} \frac{\l^2}{M^6} + \frac{\L}{2}.
\label{4Dconst}
\ee
If $w = p/\rho$ is constant, then by energy conservation on the brane, $\rho = \rho_0a^{-3(1+w)}$, and
the effective Friedmann equation (\ref{complFried}) can be expressed in terms of the sole $a$.

The effective Friedman equation contains, apart from the quadratic term in the brane energy density, a
dark radiation term proportional to $\mu$ \cite{Binetruy:1999ut,Binetruy:1999hy} and a term that behaves like
``stiff matter", proportional to $Q$ \cite{Mukherji:2002ft}. But, as pointed out in the previous section,
in our case this last term, which dominates at high curvature, has a positive contribution, hence it cannot
drive a bounce in the cosmological evolution. Nevertheless, we can allow for a negative contribution
from the dark radiation term, by letting $\mu$ be negative. There is always an ``intermediate" curvature
regime in which the negative dark radiation dominates, so that it can be responsible for a possible bounce.
Notice that the bounce would happen at a larger radial coordinate in the bulk than in the case discussed in
\cite{Mukherji:2002ft}, so that the instabilities which seem to rule out the bounce \cite{Hovdebo:2003ug}
could be avoided. In order to discuss these issues, let us simplify the setup
by considering a pure tension, spatially flat brane. The Friedman equation reduces to
\be
H^2 = \frac{\L_4}{3} + \frac{\mu}{a^4} + \frac{Q^2}{3a^6}.
\label{Fried}
\ee
Notice that, the 4D cosmological constant becomes dominant at late time, so that, in order to have
an asymptotically de Sitter universe, it must be positive. This is always true in the dS bulk, while
in the AdS case the tension has to satisfy the inequality
\be
\l > \sqrt{-\frac{3}{2} \L M^6}.
\label{tenslimit}
\ee

\begin{figure}[ht]
\begin{center}
\epsfig{file=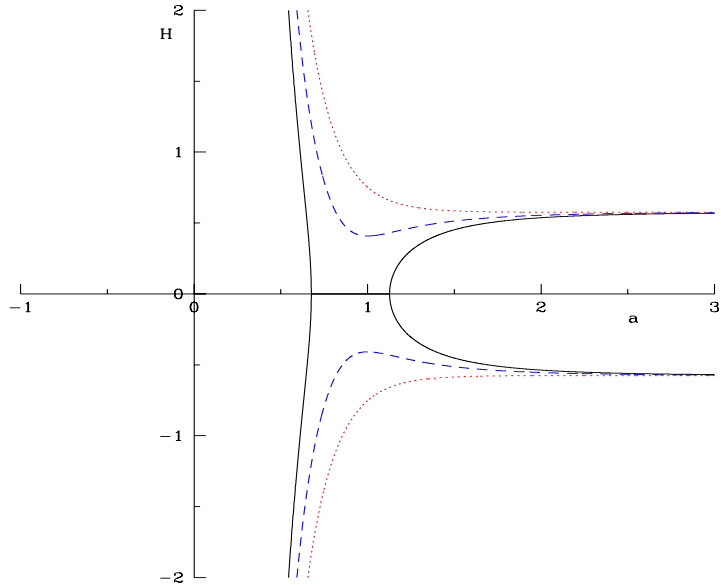,width=11cm,height=9cm}
\caption{The Hubble parameter $H$ as a function of the scale factor $a$. The three colors represent decreasing
values of $\frac{R_{KR}}{\ell_4}$: $15.81$ (red, dotted), $1.41$ (blue, dashed), $0.70$ (black, solid.}
\label{H_vs_a}
\end{center}
\end{figure}

A bounce occurs if the scale factor $a(t)$ reaches a minimum: $\dot{a}=0$, $\ddot{a}>0$.
The first condition can only be met when $\mu < 0$. The behaviour of $H$ as a function of $a$ is depicted in Fig. \ref{H_vs_a}
for different values of the parameter
\be
\frac{Q^2\sqrt{\L_4}}{2(-\mu)^{3/2}} \equiv \frac{R_{KR}}{\ell_4},
\label{alpha}
\ee
which corresponds to the ratio between the 4D de Sitter curvature radius and the characteristic length of the
Kalb-Ramond black hole obtained by the charge to mass ratio.

There are two values of $a$ for which $H = 0$, but, of course, the only one compatible with a late time de Sitter
evolution is the one located at the largest value of $a$.
As the scale factor shrinks, the Hubble parameter follow the negative branch of the plot backwards to zero,
then it start to grow again, following the positive branch. Note that this heuristic observations prove that
the point at which the Hubble parameter goes to zero (for the largest branch) is actually a minimum of
the scale factor\footnote{Quite interestingly,
the smallest branch that starts from $a=0$ can mimic the evolution of a closed universe, in which the scale
factor reaches a maximum and then starts to decrease, even if the space geometry of the brane is assumed to be flat.
But we will not discuss this issue any further.}.
Quantitatively, the zeroes of the equation (\ref{Fried}) can be obtained by solving the equation:
\be
x^3 + \frac{3\mu}{\L_4}x + \frac{Q^2}{\L_4} = 0.
\label{bounceeq}
\ee
We require $\L_4 > 0$ to have a well behaved late time evolution and $\mu < 0$ for the bounce to occur.
We find that there are three real solutions to eq. (\ref{bounceeq}) when
\be
0 < \L_4 < -\frac{4\mu^3}{Q^4} \Leftrightarrow R_{KR} < \ell_4,
\label{bounceint}
\ee
The solutions can again be expressed in terms of the Chebyshev radicals given by eq. (\ref{horsol2})
with $\L$ replaced by $2\L_4$.
The second solution is negative so it has to be discarded. We are only interested
in the larger of the two solutions $x_1$. So we conclude that a bounce can actually
occur at the scale factor
\be
a_b = \sqrt{2} \left(-\frac{\mu}{\L_4}\right)^{1/4} \sqrt{\cos \left[ \frac{1}{3}
\arccos \left( -\frac{Q^2\sqrt{\L_4}}{2(-\mu)^{3/2}}\right)\right]}.
\label{abounce}
\ee

Next we need to discuss the sign of the second derivative of $a$, to prove that $a_b$ is actually a minimum as expected.
Taking the derivative of (\ref{Fried}) we find:
\be
\frac{\ddot{a}}{a} = -\frac{2}{3} \frac{Q^2}{a^6} - \frac{\mu}{a^4} + \frac{\L_4}{3},
\label{addoteq}
\ee
The sign of $\ddot{a}$ at the bounce can be deduced by studying the function:
\bea
\left. \frac{3a^5}{\L_4}\ddot{a} \right|_{a=a_b} = a_b^6 - \frac{3\mu}{\L_4}a_b^2 - \frac{2Q^2}{\L_4} =
-\frac{6\mu}{\L_4}a_b^2 - \frac{3Q^2}{\L_4} \nn \\
= 12 \left( -\frac{\mu}{\L_4} \right)^{3/2} \cos \left[ \frac{1}{3}
\arccos \left( -\frac{Q^2\sqrt{\L_4}}{2(-\mu)^{3/2}}\right)\right] - \frac{3Q^2}{\L_4},
\label{IIdiseq1}
\eea
where the second equality comes from using (\ref{bounceeq}). Now, observing that $\cos[\ldots] > 1/2$ and using
inequality (\ref{bounceint}) we get
\be
\left. \frac{a^5}{\L_4}\ddot{a} \right|_{a=a_b} > 2 \left( -\frac{\mu}{\L_4} \right)^{3/2} - \frac{Q^2}{\L_4}
> 2 \left( \frac{\L_4 Q^4}{4}\right)\frac{1}{(\L_4)^{3/2}} - \frac{Q^2}{\L_4}= 0,
\label{IIdiseq2}
\ee
which proves that there is actually a bounce at $a=a_b$.

The crucial requirement for a bounce to be acceptable is that it occurs before the brane crosses the black hole
horizon. So we have to check that there exist a non-empty region of the parameter space in which $a_b > R_+$ or $a_b > R_0$.
\begin{center}
\begin{figure}
 \begin{minipage}[b]{8cm}
   \includegraphics[width=7cm]{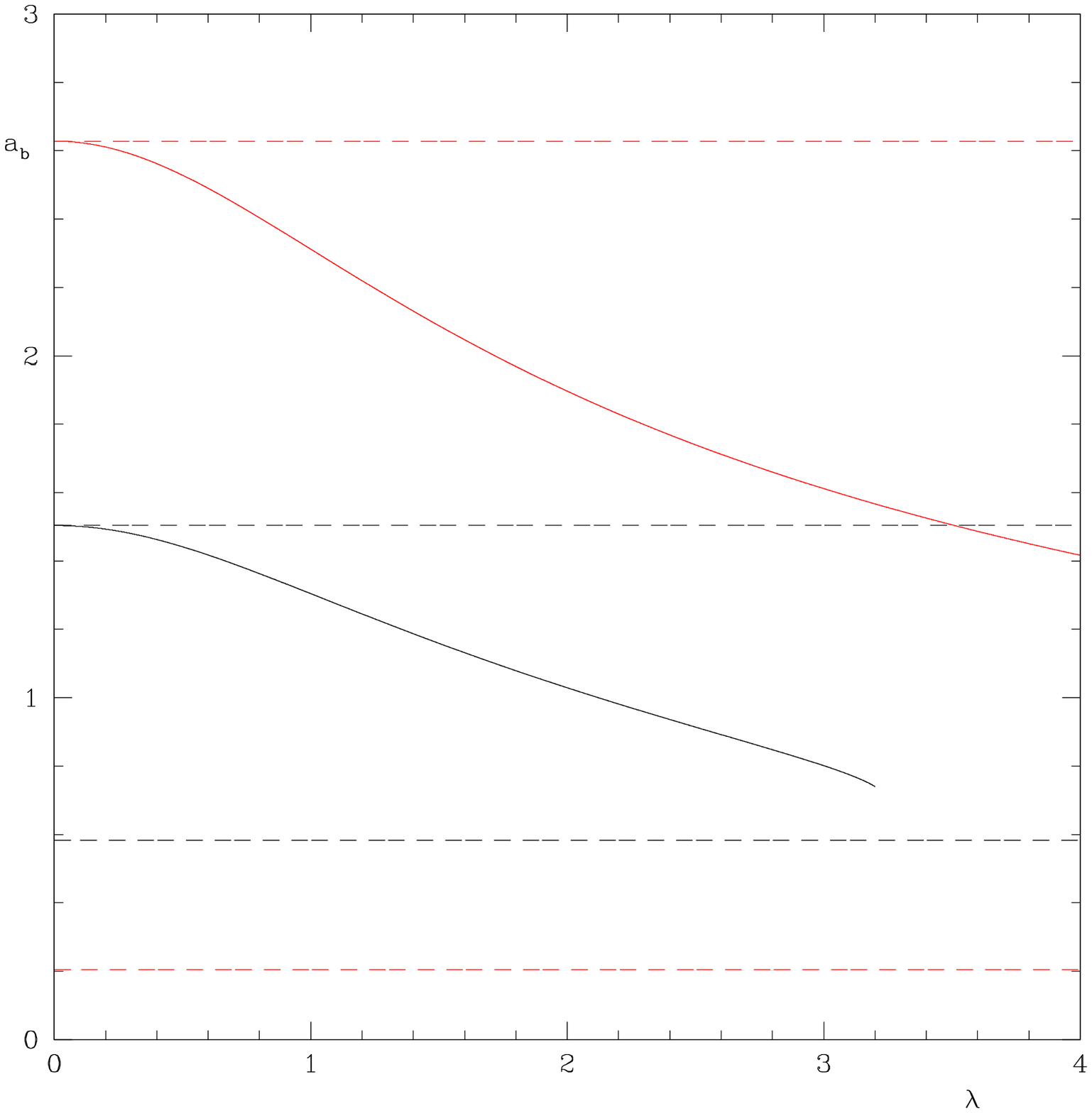}
 \end{minipage}
 \begin{minipage}[b]{8cm}
   \includegraphics[width=7cm]{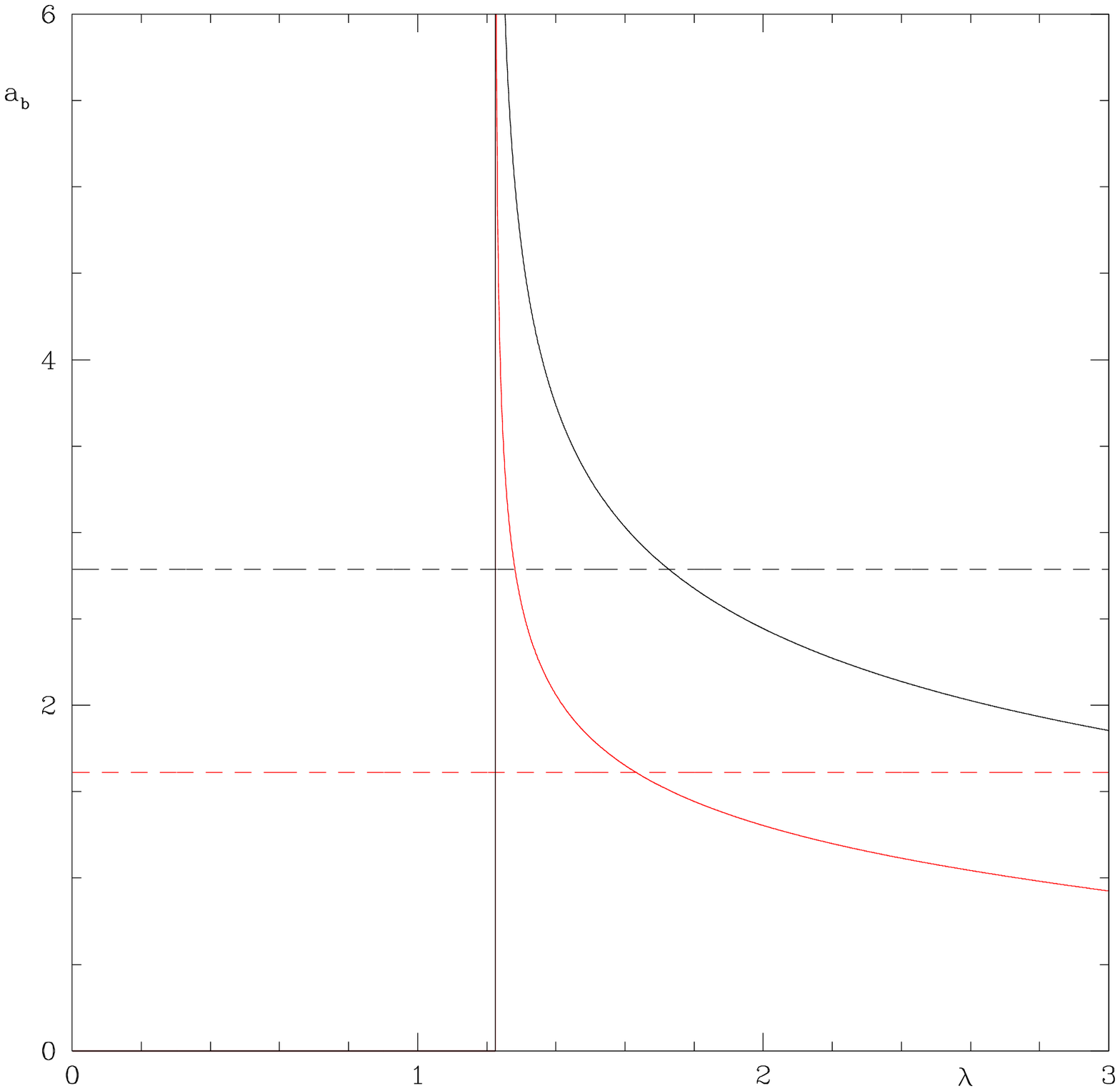}
 \end{minipage}
\caption{Plot of $a_b$ as a function of the tension $\l$ for different values of $\frac{R_{KR}}{\ell_5}$: $0.5$
(black, darker), $0.02$ (red, lighter). Dashed curves of the same color represent the position of the horizons for the same value of $\frac{R_{KR}}{\ell_5}$. The case with 2 horizons is on the left, the case with one horizon is on the right.}
\label{horcomp}
\end{figure}
\end{center}

We discuss briefly the case in which the bulk black hole has two horizons, i.e. case 2 of the
previous section, so that the physical parameters satisfy the inequality (\ref{horint2}).
As we will see, this case has to be discarded. In fact, Fig. \ref{horcomp} (left) shows a plot of $a_b$
as a function of the brane tension $\l$, and the position of the two horizons $R_+$ and $R_-$.
We see that the position of the bounce is always inside the external horizon, and coincides with it for the tensionless brane.
Analytically, it is easy to see that $a_b$  is a decreasing function of $\l$, and, since $\L_4 \rightarrow \L/2$ as
$\l \rightarrow 0$, we readily get $a_b \rightarrow R_+$ in the tensionless limit. So, this case is expected
to have an instability similar to the one discussed in \cite{Hovdebo:2003ug} (even though in that case
the brane actually bounces after crossing {\it both} horizons).

Next, we discuss the most interesting case of a negative bulk cosmological constant and a single horizon,
case 3 of the previous section. Fig. \ref{horcomp} (right) shows $a_b(\l)$, and the corresponding
value of $R_0$. In contrast to what happens with two horizons, there is a region, when the brane tension is
close to the critical value expressed in eq. (\ref{tenslimit}), in which the bounce radius is greater
than the horizon position, so that the entire evolution of the
brane lies in the physically viable region outside the horizon. This feature is quite general, and the reason is
easy to understand, since we can see from (\ref{abounce}) that $a_b \ra \infty$ as $\L_4 \ra 0$.
Analytically, we have to solve the inequality $a_b>R_0$. After some algebra, we find
that $\L_4$ has to satisfy the following inequalities:
\bea
\L_4 < -\frac{\L}{2}\frac{\frac{3}{2}C_{1/3}\left(2\frac{Q^2\sqrt{-\L}}{(-2\mu)^{3/2}}\right) -
\frac{Q^2\sqrt{-\L}}{(-2\mu)^{3/2}}}
{\frac{3}{2}C_{1/3}\left(2\frac{Q^2\sqrt{-\L}}{(-2\mu)^{3/2}}\right) + \frac{Q^2\sqrt{-\L}}{(-2\mu)^{3/2}}}
&~{\rm for}~& \frac{Q^2\sqrt{-\L}}{(-2\mu)^{3/2}}<5, \\
\L_4 < -\frac{\L}{4} C_{1/3}^{-2}(2\frac{Q^2\sqrt{-\L}}{(-2\mu)^{3/2}}) &~{\rm for}~& \frac{Q^2\sqrt{-\L}}{(-2\mu)^{3/2}}>5,
\label{4DLint}
\eea
where the Chebyshev polynomials have been defined in (\ref{Chebys}). So there is always an allowed value of the
brane tension for which the brane evolution lies entirely outside the horizon.

\begin{figure}[ht]
\begin{center}
\epsfig{file=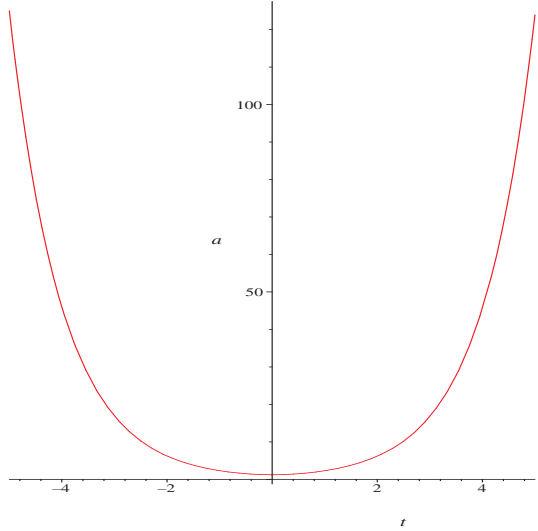,width=7cm,height=7cm}
\caption{Plot of the numerical solution of $a(t)$. The initial conditions used for the numerical integration are
$a(t=-5)=125$, $\dot{a}(t=-5)=-125$.}
\label{a_plot}
\end{center}
\end{figure}

Finally, to get the full dynamical evolution from a 4D perspective, we need to solve the effective Raychaudhuri
equation (\ref{addoteq}). It can not be solved analytically, but numerical solution confirms that the scale
factor undergoes a bounce from a contracting phase to an expanding one, as illustrated in Fig. \ref{a_plot}.

\section{Conclusions and outlook}
\label{Concl}

In this paper we presented a braneworld model in which the cosmological evolution of the brane is non-singular, and
the brane universe bounces smoothly from a phase of contraction to a subsequent expanding phase.
The cosmological evolution on the brane is induced by its movement through a static bulk AdS black
hole supported by a non-trivial Kalb-Ramond antisymmetric 2-form.
The solution is similar to the Reissner-Nordstr{\o}m-AdS solution
presented previously in the literature \cite{Mukherji:2002ft}, but in our model the bounce is induced
by a negative dark radiation term sourced by the black hole mass term. In order for this to be possible,
the integration constant $\mu$ which appears in the metric solution (\ref{backgroundsol}),
and which is proportional to the mass of the black hole, must be negative.
Black holes with negative mass have been considered in the literature \cite{Mann:1997jb,Birmingham:1998nr}.
This does not lead to a naked singularity, as we have shown in section \ref{Horizons}.
But it may create some difficulties related with the
overall stability of the solution (and possibly to its physical interpretation).
Stability analysis of charged black holes in 5 dimensions has proven to be a very difficult task,
and there appears to be no definitive answer for the standard Reissner-Nordstr{\o}m-(A)dS
black hole \cite{Kodama:2003kk}. More recently, it has been proved \cite{Birmingham:2007yv}
that a class of topological black holes with negative mass is stable in every dimension, which
can be regarded as hint of the good behaviour of 5-dimensonal negative mass black holes against perturbations.
The thermodynamical interpretation of the black hole solution requires
the temperature of the black hole to be positive. The temperature can be related to the
value of the derivative of $f$ at the horizon, so the latter must be positive in order to have
an acceptable thermodynamical behaviour. In our solution this is always the case,
whatever the value of $\mu$. Stability and thermodynamics of the bulk solution presented in this paper
are worth further investigation, which we plan to do in a forthcoming paper.

Future investigation also should extend the study to more general settings. For example, we
have only considered the case of a spatially flat, pure tension brane. While the curvature
term is not expected to change dramatically the picture, it is possible that the presence of matter
could spoil the bouncing behaviour. In fact, the bounce is driven by the negative dark radiation term, so a
sufficiently large amount of positive ``ordinary'' radiation could possibly compensate for the
negative energy term, thus making the singularity appear again. The study of the case of a
non-empty brane is also important to test the model against the observational constraints on the early
universe evolution, such as nucleosynthesis. In practice, it is a nice feature that the bounce occurs at a large scale,
but it should not be ``too large'', so that the universe would have enough time to undergo its standard evolution.

Another interesting development would be to consider a charged brane. The string theory embedding of the
braneworld scenario would require the interpretation of the brane as a $D$-brane on which the open strings,
which represent gauge particles, are confined. Of course, a full string theory derivation of braneworld models
is still lacking. Nevertheless, a $D$-brane is naturally coupled to the Kalb-Ramond 2-form via, for example, a
Dirac-Born-Infeld action, which can be implemented also in the present model. But in this case the
ansatz (\ref{HAnsatz}) is not expected to hold, so the calculation becomes much more complicated.

Bouncing cosmologies have been proposed in various contexts \cite{Gasperini:1996fv,Giovannini:1998xw,DeRisi:2001ed,
Gasperini:1996fu,Brandenberger:1998zs,Easson:1999xw,Foffa:1999dv,Cartier:1999vk,Ashtekar:2006rx,Ashtekar:2006wn,
DeRisi:2007gp,Khoury:2001bz,Steinhardt:2001st,Germani:2006pf} as an alternative to inflation.
They have of course the remarkable
feature of avoiding the initial singularity, but also provide for an alternative, but still perfectly
acceptable, solution of the kinematic problems of the standard cosmological model, without invoking an
unknown inflaton. Nevertheless, scalar fluctuations observed  by the WMAP and SDSS \cite{Spergel:2006hy}
experiments seem to favour a nearly scale-invariant spectrum, which is in agreement
with the prediction from chaotic inflation, but very difficult
to obtain in a bouncing model. So, as a second step, it would be essential to
study the behaviour of the scalar (and tensor) perturbations.

\section*{Acknowledgements}

It is a pleasure to thank Valerio Bozza, Olindo Corradini, Maurizio Gasperini, Francisco Lobo, Roy Maartens,
Sanjeev Sehara and David Wands for helpful discussions and comments on the manuscript.
GDR is supported by INFN.

\end{document}